\documentclass[twocolumn,prl]{revtex4}
\usepackage{graphicx}    
\usepackage{dcolumn}     
\usepackage{bm}          
\newcommand{\Vec}[1]{\mbox{\boldmath$#1$}}
\begin{document}

\title{Possible spin-triplet $f$-wave pairing due to 
disconnected Fermi surfaces in Na$_x$CoO$_2\cdot y$H$_2$O
}
\author{
Kazuhiko Kuroki
}
\affiliation{
Department of Applied Physics and
Chemistry, The University of Electro-Communications,
Chofu, Tokyo 182-8585, Japan}

\author{Yukio Tanaka}
\affiliation{Department of Applied Physics,
Nagoya University, Nagoya, 464-8603, Japan}

\author{Ryotaro Arita}
\affiliation{Department of Physics,
University of Tokyo, Hongo 7-3-1, Tokyo 113-0033, Japan}

\date{\today}
\begin{abstract}
We propose that spin-triplet pairing mechanism 
due to disconnected Fermi surfaces proposed in 
our previous study [Phys. Rev. B {\bf 63} 174507 (2001)] 
may be at work in a recently discovered superconductor 
Na$_x$CoO$_2$ $\cdot y$ H$_2$O.
We introduce a single band effective model that takes into account the 
pocket-like Fermi surfaces along with the van Hove singularity 
near the K point  found in the band calculation results.
Applying fluctuation exchange method and solving the linearized 
{\'E}liashberg equation, 
the most dominant pairing is found to have spin-triplet $f$-wave symmetry, 
where the nodes of the gap function 
do not intersect the pocket Fermi surfaces. Presence of 
finite $T_c$ is suggested in sharp 
contrast with cases when the gap nodes intersect the Fermi surface.
\end{abstract}
\pacs{PACS numbers: }
\maketitle

Recent discovery of superconductivity in Na$_x$CoO$_2\cdot y$H$_2$O
\cite{Takada} has attracted much attention. 
One reason why the material is of great interest 
is that it is a $d$-electron system having a quasi-two-dimensional 
lattice structure, which resembles the high $T_c$ cuprates.
On the other hand it differs from the cuprates in that 
the Co ions form a triangular lattice, and also there are several
bands intersecting the Fermi level, all of which are quite away 
from half-filling. Another interesting point concerning this material is 
that some experiments suggest possible occurrence of an 
unconventional superconductivity,\cite{Sakurai,Waki,Fujimoto,Ishida} 
although there are also experiments suggesting conventional pairing.
\cite{Kobayashi}
In particular, an unchanged 
Knight shift across $T_c$ found in some experiments 
suggests spin-triplet pairing,\cite{Waki} while 
absence of broken time reversal symmetry\cite{Higemoto,Uemura}
may exclude possibility of chiral $p$-wave or $d$-wave
pairing states, leaving spin triplet $f$-wave 
pairing as a good candidate.

Theoretically, although importance of multiband effects
has been proposed,\cite{Koshibae} 
number of studies have considered a single band model on a 
triangular lattice,\cite{Baskaran,Kumar,Wang,Ogata,Tanaka,Ikeda,TanaOgata}
and in particular focused on the band filling (number of electrons 
per site) $n=1.33$, 
corresponding to the nominal number of $t_{2g}$ holes ($0.67$ holes) per Co
in the case of Na$_{0.33}$CoO$_2$.
Several of these theories have indeed proposed possible occurrence 
of spin-triplet superconductivity,\cite{Tanaka,Ikeda,TanaOgata} 
and in particular $f$-wave pairing.\cite{Ikeda,TanaOgata}

\begin{figure}[htb]
\begin{center}
\scalebox{0.8}{
\includegraphics[width=10cm,clip]{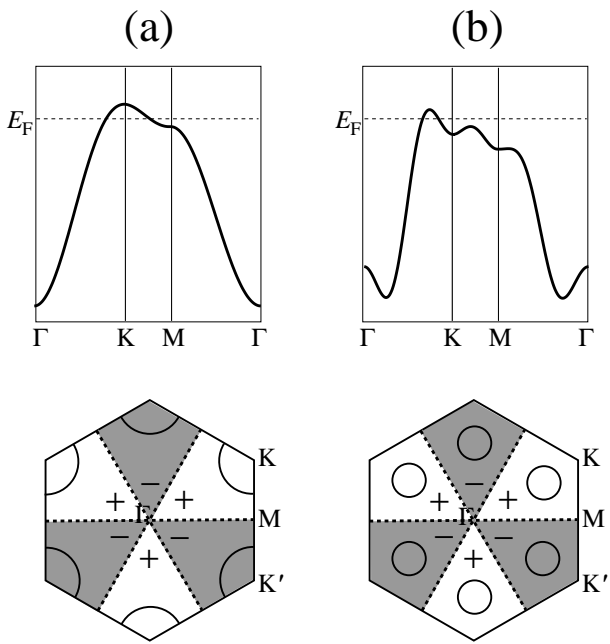}}
\caption{
Band dispersion (top) along with the Fermi surfaces (schematic)
for a small number of holes and the $f$-wave gap 
(bottom) are shown for a tight binding model on a triangular lattice 
with (a) only nearest neighbor hopping, and (b) when 4th nearest neighbor
hopping $t_4=0.2$ is introduced. 
$+-$ denote the sign of the gap function, while the 
dashed lines represent the nodal lines.
\label{fig1}}
\end{center}
\end{figure}

In fact, the presence of ferromagnetic spin fluctuations observed 
experimentally\cite{Ishida,Kobayashi} 
indirectly support the spin-triplet pairing scenario.
Couple of years ago, however, two of the present authors showed that 
$T_c$ of spin triplet superconductivity due to ferromagnetic 
spin fluctuations should  in general be very very low (much lower than 
the $T_c$ in Na$_x$CoO$_2$) or does not even exist.\cite{AKA}
A similar conclusion has been reached by Monthoux and Lonzarich.\cite{ML}
This is because the triplet pairing interaction in the spin-fluctuation 
scenario is proportional to 
$\chi/2$ (in contrast with $3\chi/2$ for the singlet case), 
where $\chi$ is the spin susceptibility, while the 
effective interaction (denoted as $V^{(1)}$ later on in the formulation part) 
that enters the normal self energy is proportional 
to $3\chi/2$, so that the depairing effect due to the normal self-energy 
overpowers the pairing effect.
Succeedingly, we have then shown that 
this difficulty for spin-fluctuation-mediated 
triplet pairing may be eased in systems having 
`disconnected Fermi surfaces', where the nodal lines of the odd-parity gap 
can run {\it in between} the Fermi surfaces to open up a 
full gap {\it on the Fermi surfaces}.\cite{KA} 
Since one of the reasons that degrade 
pairing with high angular momentum is the presence of 
gap nodes on the Fermi surface, opening a full gap 
may result in an enhanced pairing.\cite{KA,KA2}
As an example, we have considered 
the Hubbard model on a triangular lattice, where the Fermi surface
becomes disconnected into two pieces centered around the K point and the
K$'$ point (see Fig.\ref{fig1}(a)) 
when there are a small number of holes (or 
electrons, depending on the sign of the hopping integral).
Using fluctuation exchange approximation and solving the linearized 
{\'E}liashberg equation, we have shown that a finite $T_c$ for spin-triplet 
$f$-wave pairing may be obtained at 
a $T_c$ of order $0.001t$ ($t$ is the hopping integral), 
which is much higher than the $T_c$ obtained by the same method, 
if any, for systems having enhanced ferromagnetic
spin fluctuations but with connected Fermi surfaces.
\begin{figure}[htb]
\begin{center}
\scalebox{0.8}{
\includegraphics[width=10cm,clip]{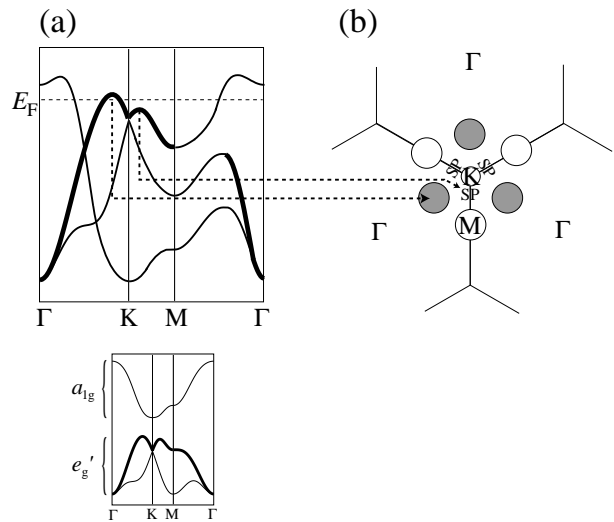}}
\caption{
(a) A schematic plot of the band structure of Na$_x$CoO$_2$ following
the calculation results of ref.\cite{Singh}.
The thick line denotes the portion of the upper part of the bands 
which has $e_g'$ character to some extent. Bottom inset:
dispersion of a three band tight binding model, where large 
$a_{1g}$-$e_g'$ level offset as well as third nearest neighbor hoppings 
between same orbitals is introduced in addition to the hoppings considered in 
ref.\cite{Koshibae}.
(b) Brillioun zone is shown in the extended zone scheme. 
The points denoted as SP are saddle points since the energy of the thick band 
at SP is lower than those around the gray circles, while it is 
higher than those around the open circles. 
\label{fig2}}
\end{center}
\end{figure}

If we look at the  band calculation results of Na$_{x}$CoO$_2$\cite{Singh} 
from this point of view, we find that the pocket-like 
Fermi surfaces near the K and K$'$ points (as in the bottom figure of 
Fig.\ref{fig1}(b)), originating from the band having $e_g'$ character
to some extent in the notation of ref.\cite{Singh} 
(denoted by the thick line in Fig.\ref{fig2}(a)), 
are disconnected in a similar sense as in 
the triangular lattice, namely the nodes of the $f$-wave gap 
do not intersect the Fermi surfaces 
(although they do intersect the large Fermi surface
around the $\Gamma$ point originating from the bands 
having $a_{1g}$ character). 
This viewpoint has motivated us to 
investigate possible occurrence of $f$-wave superconductitivy 
due to these pocket-like Fermi surfaces.

Another reason to focus on the pocket Fermi surfaces is the 
presence of van Hove singularity (vHS) near the K point.
Namely, as can be seen from the band structure 
shown in Fig.\ref{fig2}, there exist saddle points at points denoted as SP,
where the density of states (DOS) takes a large value. Since this large 
DOS lies close to the Fermi level\cite{Singh}, 
it is likely that the band structure around 
the K and K$'$ points strongly affects the low energy properties 
of this material. In particular, ferromagnetic 
spin fluctuations may arise due to this high DOS near the Fermi level.

To our knowledge, 
the effect of the pocket Fermi surfaces and the vHS has not been
properly considered in previous studies.
Although it is necessary to consider a multiband model in order to 
strictly take into account these effects, 
we believe that a good starting point for the present 
purpose is to separate out the portion of the 
bands\cite{commentmb} which has $e_g'$ character to some extent 
(we will call these bands `$e_g'$ bands' for simplicity), 
and in particular focus only on the upper $e_g'$ band,
which contributes to the formation of the pocket Fermi surfaces 
and to the large DOS at the vHS, while neglecting 
the lower $e_g'$ band, which has only small DOS 
near the Fermi level due to the linear dispersion at the band top. 
Although this portion of the band, 
shown by the thick curves in Fig.\ref{fig2}(a),
is disconnected between M and $\Gamma$ points due to $a_{1g}$-$e_g'$
hybridization, it originally comes from a single band having 
$e_g'$ character, as can be understood from the 
inset of Fig.\ref{fig2}(a), where a tight binding band dispersion is given 
for a case when large $a_{1g}$-$e_g'$ level offset is introduced.
In fact, we have found that the thick portion of the band 
in Fig.\ref{fig2}(a) (apart from the 
missing part between M and $\Gamma$ points) 
can be roughly reproduced by a single band tightbinding 
model on a triangular lattice 
with hopping integrals up to 4th nearest neighbor.
Namely, the dispersion of the tight binding model 
on an isotropic triangular lattice with only nearest neighbor hoppings 
takes its maximum at the K point (Fig.\ref{fig1}(a)), while 
the band top moves towards the $\Gamma$ point 
when 4th neighbor hopping is introduced  (Fig.\ref{fig1}(b)),
resulting in a pocket like Fermi surface that lies between the 
K and $\Gamma$ points when small amount of holes are present.
Moreover, the band structure around the K point resembles 
that of the actual material (compare Fig.\ref{fig1}(b) and the 
thick curve in Fig.\ref{fig2}), so the vHS at points SP in 
Fig.\ref{fig2} is also reproduced by this effective model.
As for the electron-electron interaction, here we consider 
the Hubbard model, where we take into account only the on-site repulsion. 

In standard notations, the Hamiltonian considered in the present study 
is given in the form,
\[
H=\sum_{{ij},\sigma}\left(t_{ij}c^\dagger_{i\sigma}c_{j\sigma}+{\rm h.c.}
\right)+U\sum_i n_{i\uparrow} n_{i\downarrow}
\]
where $t_{ij}=t_1$, $t_2$, $t_3$, and $t_4$ for the nearest, 
2nd nearest, 3rd nearest and 4th nearest neighbors, respectively.
We take $t_1=-1$ throughout the study.
We apply the fluctuation exchange (FLEX) approximation
\cite{Bickers,Dahm,Grabowski}, where (i) Dyson's equation is 
solved to obtain the renormalized Green's function $G(k)$,
where $k\equiv({\bf k},i\epsilon_n)$ denotes the 2D wave-vectors and 
the Matsubara frequencies,
(ii) the effective electron-electron interaction $V^{(1)}(q)$ 
is calculated by collecting RPA-type bubbles and ladder diagrams consisting
of the renormalized Green's function, namely, 
by summing up powers of the irreducible susceptibility 
$\chi_{\rm irr}(q)\equiv -\frac{1}{N}\sum_k G(k+q)G(k)$ 
($N$:number of $k$-point meshes),
(iii) the self energy is obtained as 
$\Sigma(k)\equiv\frac{1}{N}\sum_{q} G(k-q)V^{(1)}(q)$, 
which is substituted into Dyson's equation in (i), 
and the self-consistent loops are repeated until convergence is attained.
Throughout the study, we take up to $64\times 64$ $k$-point meshes and 
the Matsubara frequencies $\epsilon_n$ from 
$-(2N_c-1)\pi T$ to $(2N_c-1)\pi T$ with $N_c$ up to 65536 in order to
ensure convergence at low temperatures.\cite{commentconv} 

To obtain $T_c$, we solve the linearized 
{\'E}liashberg equation for the gap function $\phi(k)$, 
\begin{eqnarray*}
\lambda\phi(k)&=&-\frac{T}{N}
\sum_{k'}
\phi(k')|G(k')|^2 V^{(2)}(k-k'),
\label{eliash}
\end{eqnarray*}
where the pairing interaction $V^{(2)}$ is given as 
$
V_t^{(2)}(q)=-\frac{1}{2} \left[ \frac{U^2\chi_{\rm irr}(q)}
{1-U\chi_{\rm irr}(q)} \right]
-\frac{1}{2} \left[ \frac{U^2\chi_{\rm irr}(q)}{1+U\chi_{\rm irr}(q)} \right] 
$
for triplet pairing. (The tendency toward singlet pairing is 
weak in the present case, so we focus only on the triplet pairing.)
$T_c$ is the temperature at which the maximum eigenvalue 
$\lambda$ reaches unity.

\begin{figure}
\begin{center}
\scalebox{0.8}{
\includegraphics[width=10cm,clip]{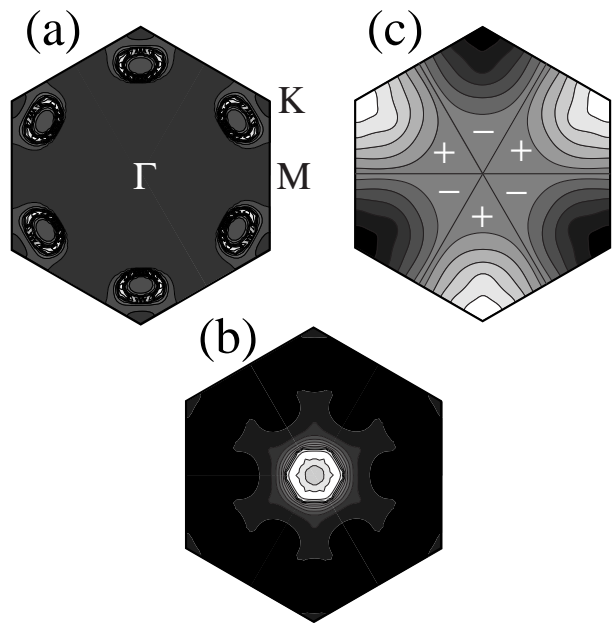}}
\caption{
Contour plots of (a)$|G(\Vec{k},i\pi k_BT)|^2$, (b)$\chi(\Vec{k},0)$, 
(c) $\phi(\Vec{k},i\pi k_BT)$ for $(t_2,t_3,t_4)=(0,0,0.2)$, $U=4.5$, 
$n_h=0.2$ and $T=0.01$ .
\label{fig3}}
\end{center}
\end{figure}

We now move onto the results. In Fig.\ref{fig3}, we present 
contour plots of $|G(\Vec{k},i\pi k_BT)|^2$, $\chi(\Vec{k},0)$, 
$\phi(\Vec{k},i\pi k_BT)$ for $(t_2,t_3,t_4)=(0,0,0.2)$ 
(the band dispersion for $U=0$ for this choice of hoppings is shown in 
the upper panel of Fig.\ref{fig1}(b)), $U=4.5$, 
number of holes per site $n_h=0.2$ and $T=0.01$. Note that we do not focus on 
$n_h\sim 0.65$ as in previous theories because the number of holes in the 
$e_g'$ band is small.
It can be seen that pocket-like Fermi surfaces, 
represented by the ridge of $|G|^2$, lie
between the K and $\Gamma$ points. Moreover, the presence of saddle points 
in $|G|^2$ at points corresponding to SP in Fig.\ref{fig2} 
indicates the presence of vHS just below 
the Fermi level as expected. 
The spin susceptibility takes large values around 
the $\Gamma$ point indicating presence of ferromagnetic spin
fluctuations. The ferromagnetic fluctuation originates 
from the large density of states near the Fermi level.
The gap function having the largest 
eigenvalue has spin-triplet $f$-wave symmetry, where the nodes of the gap 
do not intersect the Fermi surfaces. 

\begin{figure}[htb]
\begin{center}
\scalebox{0.9}{
\includegraphics[width=10cm,clip]{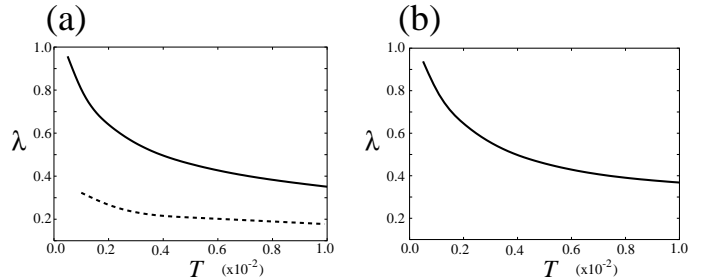}}
\caption{
$\lambda$ plotted as functions of $T$ for (a) $(t_2,t_3,t_4)=(0,0,0.2)$ (solid
 curve), $(t_2,t_3,t_4)=(-0.25,0,0)$ (dashed curve), 
and (b) $(t_2,t_3,t_4)=(-0.05,-0.1,0.15)$. $U=4.5$ and $n_h=0.2$ in all cases.
\label{fig4}}
\end{center}
\end{figure}

\begin{figure}[htb]
\begin{center}
\scalebox{0.8}{
\includegraphics[width=10cm,clip]{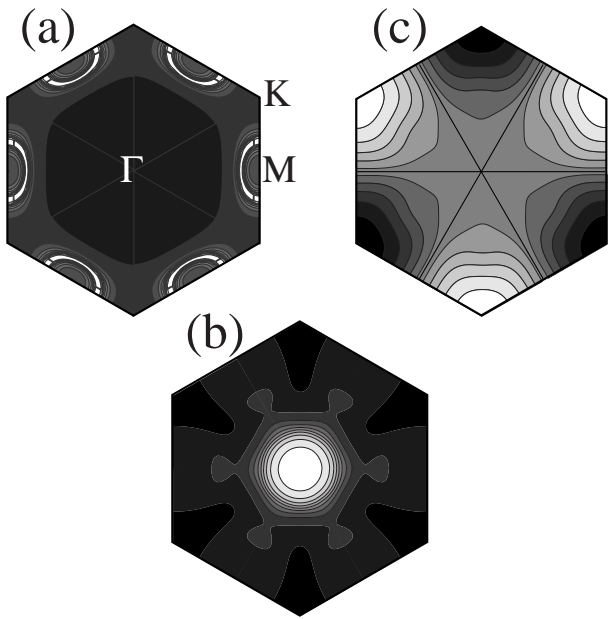}}
\caption{
Contour plots similar to Fig.\protect\ref{fig3} for 
$(t_2,t_3,t_4)=(-0.25,0,0)$, $U=4.5$, $n_h=0.2$, and $T=0.01$.
\label{fig5}}
\end{center}
\end{figure}

In Fig.\ref{fig4}(a), we show $\lambda$ as a function of temperature
(solid curve).
As can be seen, $\lambda$ tends to unity at low temperatures.
As we have mentioned in the introductory part, were it for a 
system with connected Fermi surfaces, $T_c$ of spin triplet 
superconductivity due to ferromagnetic spin fluctuations
(estimated by the present method) would be, if any, too low to be detected
in a feasible calculation. Thus we attribute the present result 
to the disconnectivity of the Fermi surface and the 
absence of gap nodes on the Fermi surfaces.
To show that large values of $\lambda$  is not restricted to a special 
choice of the hopping integral values, we show in Fig.\ref{fig4}(b) 
a result for another choice of parameter set 
which also roughly reproduces the 
$e_g'$ band structure. 
It can be seen that $\lambda$ again tends to unity at 
low temperatures.

In order to reinforce our argument, we finally consider a case where the 
$f$-wave pairing dominates over other pairing symmetries, but the nodes of the 
gap intersect the Fermi surfaces. In Fig.\ref{fig5}, we present 
a plot similar to Fig.\ref{fig3} for 
$(t_2,t_3,t_4)=(-0.25,0,0)$, $U=4.5$, $n_h=0.2$, and $T=0.01$.
We have chosen these parameter values so that the 
spin susceptibility is ferromagnetic and 
has roughly similar maximum values as in the case considered above. 
It can be seen that the 
nodes of the $f$-wave gap now intersect the Fermi surfaces.
As a result, $\lambda$ becomes much smaller as seen in Fig.\ref{fig4}
(dashed curve).\cite{commenttopt} From this result, we can also say 
that the enhancement of triplet pairing obtained in the previous cases 
(solid lines in Fig.\ref{fig4}(a)(b)) is 
not simply due to large DOS at the Fermi level 
in the BCS sense because DOS near the 
Fermi level is also large in the present case.

Now let us comment on the relation between the present theory and
experiments. Several experiments suggest presence 
of gap nodes,\cite{Fujimoto,Ishida}
which may seem to contradict with the present scenario.
In the actual Na$_x$CoO$_2$, however,
there is also the large Fermi surface around
the $\Gamma$ point, which is not taken into account in the present study.
Then we can propose a possible scenario where the `active' Fermi surfaces for 
superconductivity are the pockets, while the large Fermi surface is 
a 'passive' one, in which superconductivity is 
induced by the `active' band via interband interactions.\cite{MO}
Since the $f$-wave gap intersects the 
large Fermi surface, this can account for the experimental results 
suggesting the existence of gap nodes. In fact, it has 
recently been found that a $\mu$SR data can be well 
explained by taking this view.\cite{Uemura}

Another point to be mentioned in relation with the experiments is 
that the pocket Fermi surfaces are not observed in ARPES experiments
\cite{Hasan,Yang} up to date.
However, since these experiments are done for materials with 
relatively large Na content,
(i.e., large number of electrons in CoO layers) 
it is likely that the Fermi level lies above the $e_g'$ bands.
\cite{Karppinen} In fact, an experimental result 
that maximum $T_c$ is reached only when the content of 
Na decreases sufficiently\cite{Schaak} can be considered as 
an indirect support for scenario in which the $e_g'$ 
band plays an important role for superconductivity.

Theoretically, there are some remaining problems. 
In the FLEX method adopted here,
vertex corrections are not taken into account. 
Then, one may wonder whether the present conclusion 
holds even if vertex corrections are considered since 
in some cases it is known that such corrections 
suppress unconventional superconductivity\cite{Schrieffer}.
However, we have recently performed dynamical cluster approximation 
study on a triangular lattice, which automatically takes into account
the vertex correction effects, and have found that the $f$-wave pairing 
susceptibility is strongly enhanced upon lowering the temperature.\cite{AKA2} 
Thus, the present conclusion should hold even 
if we take into account the vertex corrections. 

Secondly, we have neglected the off-site repulsions in our model.
The study on the effect of off-site repulsion is now underway, 
where we find that for moderate values of off-site repulsions,  
$f$-wave pairing still dominates, while for larger off-site repulsions, 
$f$-wave gives way to singlet pairings. 
In particular, we have surprisingly found 
that singlet $i$-wave (which belongs to $A_2$ symmetry and 
changes sign 12 times around the $\Gamma$ point) 
pairing, which does not break time reversal symmetry 
as contrasted with $d$-wave pairing (which belongs to $E_2$ symmetry 
and takes the form $d+id$ below $T_c$), 
may dominate due to the peculiar disconectivity of the pocket 
Fermi surfaces considered here. So there may be a close 
competition between triplet $f$-wave and singlet $i$-wave pairings, which 
might be related to the fact that 
Knight shift is found to decrease below $T_c$ 
in some cases\cite{Kobayashi} while not in other cases.\cite{Waki}
Details on this point will be published elsewhere.

%

%
K.K. acknowledges M. Tanaka and T. Nojima for discussions.
Calculation has been performed
at the facilities of the Supercomputer Center,
Institute for Solid State Physics,
University of Tokyo.
%



\begin{thebibliography}{99}
\bibitem{commentmb} Similar theoretical approaches have been adopted for 
other multiband superconductors such as Sr$_2$RuO$_4$ and MgB$_2$, 
where it is believed that some of the bands play an essential role in the
occurrence of superconductivity, while other bands are only passive bands, 
where superconductivity is induced by interband interaction effects.
\bibitem{commentconv} In the lowest temperatures, we could not 
strictly check the convergence of the results with respect to number of sites 
since a large number of Matsubara frequencies is required.
Nevertheless, the tendency that large values of $\lambda$ are 
obtained at low temperatures should in any case be correct.
\bibitem{Karppinen} This view may be further supported by a recent study
by Karppinen {\it et al.} (cond-mat/0402484),
which shows that the actual Co valence can be smaller (the Fermi level can be 
higher) than expected from the composition Na$_x$CoO$_2$ due to oxygen vacancy.
\end{thebibliography}
\end{document}